# Open top anodic $Ta_3N_5$ nanotubes for higher solar water splitting efficiency

Sabina Grigorescu [a], Seulgi So [a], Jeong Eun Yoo [a], Anca Mazare [a], Robert Hahn [a], Patrik Schmuki [a,b,*]

[a] *Department of Materials Science and Engineering, WW4-LKO, University of Erlangen-Nürnberg, Martensstrasse 7, D-91058 Erlangen, Germany*

[b] *Department of Chemistry, King Abdulaziz University, Jeddah, Saudi Arabia*

\* Corresponding author: e-mail address: schmuki@ww.uni-erlangen.de

Link to the published article:

http://www.sciencedirect.com/science/article/pii/S0013468615305715

**Abstract**

In the present work we grow self-organized $Ta_2O_5$ nanotube layers in a $H_2SO_4$/$NH_4F$ electrolyte at various elevated temperatures. Under optimized conditions we obtain 4 µm long nanotubes that are well adherent to the substrate and can be grown very homogenous over large surface areas. Moreover, the key advantage of this approach is that an open top morphology (initiation layer free) is obtained. After a suitable conversion treatment in $NH_3$ atmosphere $Ta_3N_5$ nanotubes are obtained, that after (Co-Pi+Co(OH)$_x$) modification provide, under AM 1.5G illumination conditions, a photoelectrochemical current response of 4.7 mA cm$^{-2}$ at 1.23 V vs RHE.

## 1. Introduction

In the past years, one-dimensional (1D) semiconductor nanostructures such as nanotubes, nanowires, nanorods, have gained an increased scientific and technological interest [1-3]. These structures present an advantageous combination of high surface area, advanced transport properties and diffusion pathways (e.g. $TiO_2$ nanotubular layers), and have found wide applications in electronic, photoelectronic, electrochemical and photoelectrochemical devices [4].

One of the most promising semiconductors for PEC water splitting is $Ta_3N_5$ ( with a band gap of 2.1 eV), which can be obtained from $Ta_2O_5$ by a high temperature treatment in $NH_3$ atmosphere. Therefore, suitable 1D tantalum pentoxide nanostructures ($Ta_2O_5$) became highly investigated and can be fabricated by various methods, such as hydrothermal [5, 6], vapor phase hydrothermal [7,8] and electrochemical anodization [9]. Electrochemical anodization provides the advantage of obtaining highly aligned and directly back-contacted electrodes of nanostructured oxides on the Ta surface [10-12].

To utilize the full potential of these one-dimensional nanotubular layers as precursors for $Ta_3N_5$ structures for PEC water splitting, an open top morphology is desired as it provides un-blocked diffusion pathways of the reactants and products. However, controlling the Ta-oxide nanotube morphology by anodization is still challenging. For example, open and well adherent $Ta_2O_5$ nanotubes are reported by Cong et al. [13], in a sulfuric acid free electrolyte, but these nanostructures are short (~40 nm long). When $H_2SO_4$ is used [11, 14], much longer (~ 5 µm) and open (initiation layer free) $Ta_2O_5$ nanotube layers are obtained, but unfortunately are only poorly adherent to the substrate [11,14]. Throughout all existing literature, the key weakness is the combination of all desired features that is well adherent, long and open nanotubes.

Following, Su et al. [15] report the fast formation of well-ordered $Ta_2O_5$ nanotube arrays (by a 2 seconds galvanostatic anodization). However, despite the very fast formation, the nanotubes are closed at the top and a further chemical etching procedure as well as an ultrasonication are needed to open the tubular top surface, that also cause a partial removal of the nanotubular layer from the Ta substrate, thus leading to very small functional areas.

Additionally, we report in our previous work [16] the growth of ~4.5 µm long, well adherent, vertically aligned $Ta_2O_5$ nanotubes, though these layers present a less homogenous structure with a top initiation layer (covering the tubes), typical for anodization under the used conditions. Moreover, in a very recent report, Zhang et al. [17] and Wang et al. [18] show the formation of $Ta_2O_5$ nanotubes by a two-step anodization method at low temperatures.

In the present work, we report the formation of long, well adherent and open (initiation layer free) $Ta_2O_5$ nanotube arrays grown in very short times on large surface areas. These nanostructures can only be obtained by a double step anodization process at high temperature and with specific ramping speed. After conversion to $Ta_3N_5$, a further modification with co-catalysts is performed (in order to suppress photocorrosion and improve the overall PEC performance [6-8,13,15-21]), and we evaluated various co-catalyst as well as different combinations based on $Co(OH)_x$ and Co–Pi. Such type of nanotubes provides, due to their open nature, a significantly increased PEC water splitting performance, i.e. 4.7 mA/cm$^2$ compared to 0.4 mA/cm$^2$ at 1.23 V vs RHE for a non-open nanotubular structure.

## 2. Experimental

Prior to anodization, Ta foils (99.9%, Advent) with a thickness of 0.1 mm were grinded with SiC paper 1200, cleaned by ultrasonication in acetone and ethanol (5 minutes each) and dried in a nitrogen stream. The electrochemical anodization process took place in a two electrode system with a distance between the working electrode (Ta foil) and the counter electrode (Pt foil) of about 1 cm. The electrolyte used was a solution of 45 ml $H_2SO_4$, 2.5 ml Glycerol, 2.5 ml $H_2O$ and 0.5 g $NH_4F$. The samples were anodized at a constant voltage of 60V for 10 min at different temperatures (RT, 40°C, 60°C and 80°C) to obtain $Ta_2O_5$ NTs. During the anodization, the Ta substrate temperature was set to constant values from 40°C to 80°C using a heating plate controlled by a thermostat. Further on, a double anodization method with and without ramping was used at 80°C: namely, first a $Ta_2O_5$ nanotube layer was grown at 60V for 10 min and removed by a decal process using scotch tape. The remaining predimpled metal sheet was used as substrate in a second anodization step at 60V for 1 minute. For the ramping process, the anodization voltage was swept from 0 to 60V with 20 V/s. After anodization, the as prepared $Ta_2O_5$ nanotube layers were rinsed in ethanol and dried in $N_2$. $Ta_3N_5$ nanotubes were obtained by nitridation of $Ta_2O_5$ nanotubes in a horizontal quartz tube furnace under ~200 sccm $NH_3$ flow at 1000 $^0$C for 2 h [16].

For the $Co(OH)_x$ co-catalyst treatment, $Ta_3N_5$ nanotubes were immersed in a mixed solution of 0.2 M $CoSO_4$ and 0.1 M NaOH (1:2) for 15 min, then rinsed with deionized water and finally dried in $N_2$ [16]. For the Co-Pi treatment, the catalyst was anodically deposited on the $Ta_3N_5$ nanotube surfaces in a solution of 0.5 mM $Co(NO)_3$ in 0.1 M potassium phosphate buffer at pH=7 at 1 V for 8 min [15].

PEC water splitting measurements of the nanotube photoelectrodes were carried out in a three-electrode configuration using an Ag/AgCl reference electrode and a Pt foil counter electrode. The measured potential versus Ag/AgCl was converted to the RHE scale according to the Nernst equation ($E_{RHE} = E_{Ag/AgCl} + 0.059$ pH $+ 0.197$). 1 M KOH solution (pH=13.7) was used as the electrolyte. The photoelectrodes were illuminated with chopped AM 1.5G-simulated sunlight at 100 mW $cm^{-2}$ with different scan rates of 2, 10 and 30 mV $s^{-1}$.

A field-emission scanning electrode microscope (Hitachi FE-SEM S4800, Japan) was used for the morphological characterization of the electrodes.

The chemical composition of cobalt loaded $Ta_3N_5$ photoanodes was investigated by X-ray photoelectron spectroscopy (PHI 5600, spectrometer, USA) using AlKα monochromatized radiation. The sputter depth profiles were obtained by $Ar^+$ sputtering up to 50 nm (after sputtering, high resolution spectra were acquired). Shifts in spectra were corrected using the C1s peak at 284.4 eV. The fitting of the peaks was performed in Multipak software.

## 3. Results and Discussions

Ta-oxide tubes were grown by potentiostatic anodic oxidation of Ta in $H_2SO_4$ based electrolyte. Figure 1 shows current-transient curves obtained during anodizing of the Ta foil at 60V for 10 minutes at four different temperatures: room temperature (RT), 40, 60 and 80°C. All the curves show the three characteristic stages of nanotubes formation. In the initial stage of anodization, the current is dropping according to the high field mechanism and a compact oxide layer grows. In the second stage, initiation of irregular penetration of the compact layer with nanoscopic pores occurs and finally steady growth of self-ordered pores takes place. The initial layer on top of the ordered pores is commonly named initiation layer. In many cases this initiation layer, if not chemically dissolved during anodic growth or otherwise removed after the process, covers the tubes. In stage III, steady nanotube growth is established and is usually under diffusion control. The tubes grow continuously in length at relatively constant current densities. With increasing the anodization temperature, differences can be observed in the nanotube formation stages. The position of the characteristic wave for nanotubes formation observed in the current densities (due to higher active surface area of the nanotubes bottoms) shifts towards shorter anodization times, thus indicating that organized $Ta_2O_5$ nanotube formation takes place at an earlier stage. Moreover, the more pronounced shape at 80°C may indicate that tube formation takes place more simultaneously over the whole substrate, thus resulting in a homogenous layer thickness.

The top-view and cross-section SEM images presented in Figure 2a show the effect of anodization temperature on the morphology of the $Ta_2O_5$ nanotubes array (corresponding to the current-transient curves from Figure 1), while the relationship among length, surface morphology and adherence to the substrate of the tubes with different anodization temperatures is summarized in Figure 2b. When anodization takes place at RT (without temperature control), a top porous initiation layer is formed (indicated in the top-view SEM image), with cracks all over the anodized surface [16]. With increasing anodization temperature to 40°C, no major difference can be observed in the top morphology and the nanotubes present a similar initiation layer. A reason for this could be the Joule-heating effects at the metal-electrolyte interface due to the current which is passing through the metal-electrolyte interface during anodization [22]. However, from the SEM cross-section images it is evident that the nanotubes have already a more homogenous shape. A further increase in the anodization temperature, up to 60°C, brings the first obvious changes in the top morphology: that is, the top initiation layer becomes thinner and if the anodization is carried out for longer than 5 minutes (see Figure 2b), the obtained nanotubular layer presents a partially open surface morphology. Moreover, when the anodization process is carried out at 80°C, a significant improvement in the top morphology can be observed from early anodization times (Figure 2b), where the nanotubes show an even more open surface after only 3 minutes of anodization. SEM cross-section images reveal an increase of the average tubular layer thickness from ~4.5 μm at room temperature [16] to 6.3 μm at 40°C, 8.6 μm at 60°C and 21.8 μm at 80°C. Nevertheless, if a certain layer thickness is exceeded (higher than 10 μm), the adherence of the layer to the metal substrate is affected, resulting in a lift off of the $Ta_2O_5$ nanotube layer (upper dashed line in Figure 2b). Anodization at lower temperatures (RT or 40°C) for even longer times leads to longer tubes accompanied by lift off too, but no opening of the tubes is achieved at all.

The inset in Figure 2b depicts the growth rate of the nanotubes (layer thickness per amount of charge) versus the different temperatures, evaluated from SEM cross-sections. The growth rate drops by increasing the temperature or, in other words, the nanotubes are grown in a less efficient way at high temperatures (but overall faster). This may be attributed to the higher chemical dissolution rate of $Ta_2O_5$ at elevated temperatures, leading to a thinning of the initiation layer and leaving the tubes more open. Similar observations were reported for another valve metal oxide, namely $TiO_2$ [23].

Considering that anodization at high temperature (i.e 80°C) shows the strongest effect on diminishing of the initiation layer (compared to all the other tested temperatures), but not its removal, additional optimization of the anodization conditions at 80°C was required to obtain a completely open top morphology. As previously reported, the ordering and opening of anodic $TiO_2$ nanotubes may be improved by using "double step anodization" and ramping to reach the final voltage [23,24]. Therefore, $Ta_2O_5$ nanotubes were grown using a two-step anodization process at 80°C, with and without ramping, where pre-dimpled metal sheets (dimples remained after removal of the first anodization layer) were used as substrate for the second anodization step. Different sweep rates were tested (e.g. 5, 10 and 20V/s) and it was observed that with increasing ramping speed, a larger opening of the top surface is obtained. Therefore, we focused on a sweep rate of 20V/s (see Figure 3a) where a completely open top morphology is achieved. On the other hand, with double anodization but without ramping, the top surface remains partially closed; nevertheless, there is an improvement in morphology as compared to samples without "double step anodization", see Figure 2a 80°C. No clear differences were observed in the nanotube length, meaning that ramping is just affecting the thickness of the initiation layer on top of the tubes and not the total nanostructure morphology.

To study the impact of the open top morphology of the nanotubular layers on the solar PEC water splitting performance, the 4μm long $Ta_2O_5$ nanotubes (previously shown in Figure 3a) were converted by a high temperature treatment in $NH_3$ atmosphere to $Ta_3N_5$. While the open top morphology remains intact after nitridation, a decrease of the length [16] to ≈3μm is observed (Figure 3a). Figure 3b shows the photocurrent transients of the layers grown by double step anodization (see Figure 3a) and the reference (grown at RT conditions), loaded with $Co(OH)_x$ as co-catalyst, in 1 M KOH (pH 13.7). In comparison to the reference and the sample without ramping, an enhancement in the photoelectrochemical current response from 0.4 mAcm$^{-2}$ up to 1.5 mAcm$^{-2}$ at 1.23 V vs RHE, is observed only for the completely open nanotubular sample (with ramping). On one side, this may be ascribed to a better and more efficient loading of the co-catalyst over the more open nanotube structure and, therefore, more active surface and, on the other side, the open tubes provide better diffusion pathways (access of the electrolyte, transport of evolved $O_2$ away from the tubes) during the PEC process.

To gain more insights into the loading distribution of co-catalyst on the different nanotube layers (open vs closed), X-ray photo-electron spectroscopy (XPS) depth profile measurements were conducted for the double step anodization with ramping and for the reference from Figure 3b, and the amount of cobalt was analyzed. The results are plotted in Figure 3c. As XPS is a surface sensitive method, cobalt co-catalyst loadings on top of the nanotubes before sputtering are 5.9 at% for open tubes and 4.3 at% for covered nanotubes. However, by using depth sputtering, only minor changes in the Co loading are observed after removing 50 nm of the nanotubes tops and analyzing the composition again. This observation indicates that the co-catalyst loading in open and closed nanotubes seems to be similarly hindered. Thus suggesting that penetration/diffusion of reactants and products play a more significant role in the enhancement of the photocatalytic activity of the open tubes.

Due to the better PEC water splitting performance reported above (see Figure 3b), the open top nanotubular layers were further used to verify this effect by applying different co-catalysts loading procedures, namely $Co(OH)_x$ (immersion method), Co–Pi (electrochemical deposition method), as well as new combinations of $Co(OH)_x$ and Co–Pi. Figure 4a shows the current-potential curves of the electrodes with a scan rate of 2mV/s. Co–Pi/$Ta_3N_5$ nanotube photoanode shows better performance under low bias than the $Co(OH)_x$/$Ta_3N_5$. This result is in line with literature [13], showing that the Co-Pi decoration by electrodeposition is an effective treatment for shifting the photocurrent onset potential in the negative direction. Moreover, when the photoanode is treated with a Co-Pi co-catalyst combined with an additional immersion in $Co(OH)_x$ solution (Co-Pi+$Co(OH)_x$/$Ta_3N_5$), an enhancement in the photoelectrochemical current response up to 3.1 mAcm$^{-2}$ at 1.23 V vs RHE is observed, as compared to one step treated photoanodes (namely $Co(OH)_x$/ $Ta_3N_5$ and Co-Pi/$Ta_3N_5$). Applying the mixed co-catalyst the other way around (that is, $Co(OH)_x$+Co-Pi/$Ta_3N_5$), the combined treatment is less effective but still better than one step $Co(OH)_x$ or Co-Pi.

To make our results comparable to literature data, as reports are available for PEC performance using different sweep rates of 2mV/s [16], 10 mV/s [6,15,19,21] or 30mV/s [25], we further verified the influence of the scan rate on PEC water splitting performance using the combined co-catalysts Co-Pi+$Co(OH)_x$/$Ta_3N_5$ nanotubular photoanodes. Figure 4b shows the photocurrent transients with three different sweep rates. It is clear that, when the sweep rate increases from 2 mv/s up to 30 mV/s, an increase in the photocurrent response from 3.1 mA cm$^{-2}$ up to 4.7 mA cm$^{-2}$ at 1.23 V vs RHE is observed. A likely explanation is that, at higher scan rates, the photocorrosion is less contributing to the photocurrent response. Additional photostability measurements (data not shown) of the cobalt co-catalysts loaded Ta$_3$N$_5$ electrodes show similar trends as reported in literature [26]. In average, approx. 25% of the initial photocurrent remains after prolonged irradiation.

## 4. Summary

In the current work we present a successful method for obtaining open top morphology and high aspect ratio (~ 4µm) Ta$_2$O$_5$ nanotubes. The well-adherent nanotubular layers are grown by an electrochemical anodization process in very specific conditions, i.e at high temperature (80°C), "double step anodization" and ramping. After conversion, the Ta$_3$N$_5$ nanotubular layers yield a photoelectrochemical current response of 4.7 mA cm$^{-2}$ at 1.23 V vs RHE when a combined co-catalyst Co-Pi+Co(OH)$_x$ treatment is used. From the present results, it is clear that an open top morphology of Ta$_3$N$_5$ nanotubes leads to an improvement in the PEC water splitting performance.

## 5. Acknowledgements

The authors would like to acknowledge financial support from ERC, DFG within the SPP 1613 and the Cluster of Excellence (Engineering of Advanced Materials) of the FAU. Anja Friedrich is acknowledged for SEM.

Hydroxides as a Cocatalyst Strongly Stabilizing Photoanodes in Water Splitting, Chem. Mater. 27 (2015) 2360-2366.

**Figure captions:**

Figure 1: Current vs anodizing time profiles for different anodization temperatures (RT, 40°C, 60°C and 80°C).

Figure 2: (a) Top and cross-section SEM images for $Ta_2O_5$ nanotubes grown at 60V for 10 min with four different anodization temperatures (RT, 40, 60 and 80°C). (b) Length vs time curves for all tested anodization temperatures; the inset shows the growth rate (µm/C) vs temperature (°C).

Figure 3: (a) Top and cross-section SEM images of nanotubes grown at 80°C, on pre-dimpled substrates using a "double step anodization" method: $Ta_2O_5$ nanotubes without ramping, $Ta_2O_5$ and $Ta_3N_5$ nanotubes with ramping; (b) Photocurrent-potential curves of a closed $Ta_3N_5$ layer (reference) and open $Ta_3N_5$ nanotubes grown by two steps anodization with and without ramping. $Co(OH)_x$ was used as co-catalyst and spectra were obtained under chopped AM 1.5G simulated sunlight in 1 M KOH solution (pH=13.7) and a scan rate of 2 mV/s; (c) XPS depth profile measurements of loading distribution of the $Co(OH)_x$ co-catalyst on different nanotubes (open vs closed).

Figure 4: (a) Photocurrent-potential curves of open $Ta_3N_5$ nanotubes loaded with different co-catalysts ($Co(OH)_x$, Co-Pi, $Co(OH)_x$+Co-Pi and Co-Pi+$Co(OH)_x$) under chopped AM 1.5G simulated sunlight in 1 M KOH solution (pH=13.7) and a scan rate of 2 mV/s; (b) Photcurrent-potential curves for Co-Pi+$Co(OH)_x$/$Ta_3N_5$ electrode with different scan rates (2mV/s, 10 mV/s, 30 mV/s).

**Figure 1**

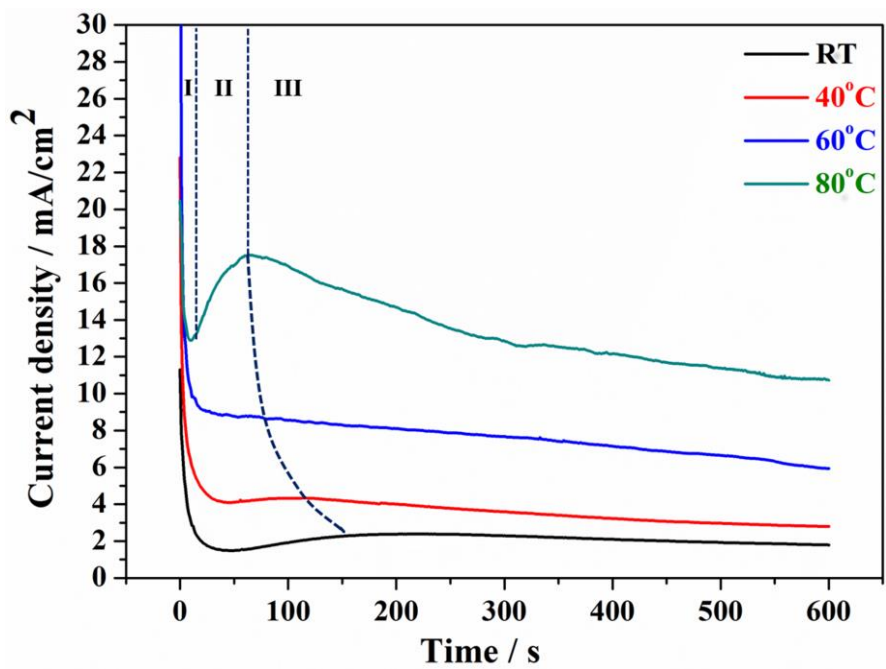

**Figure 2**

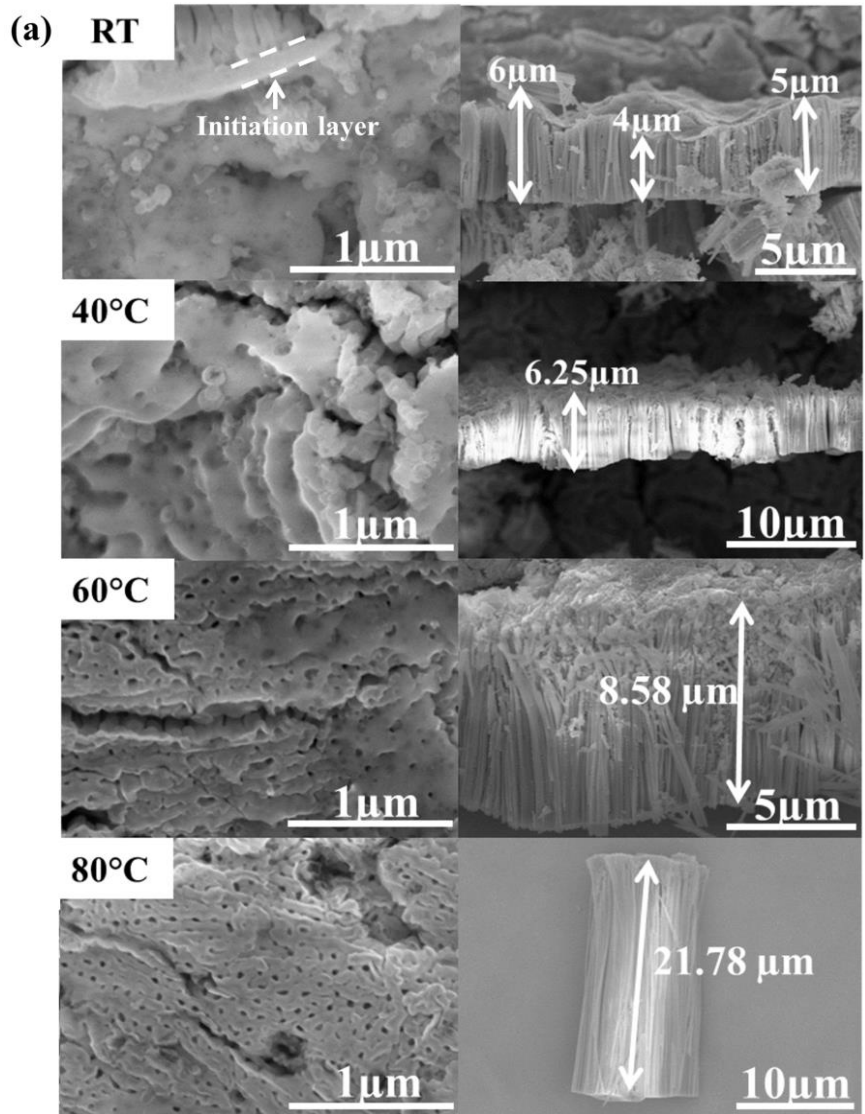

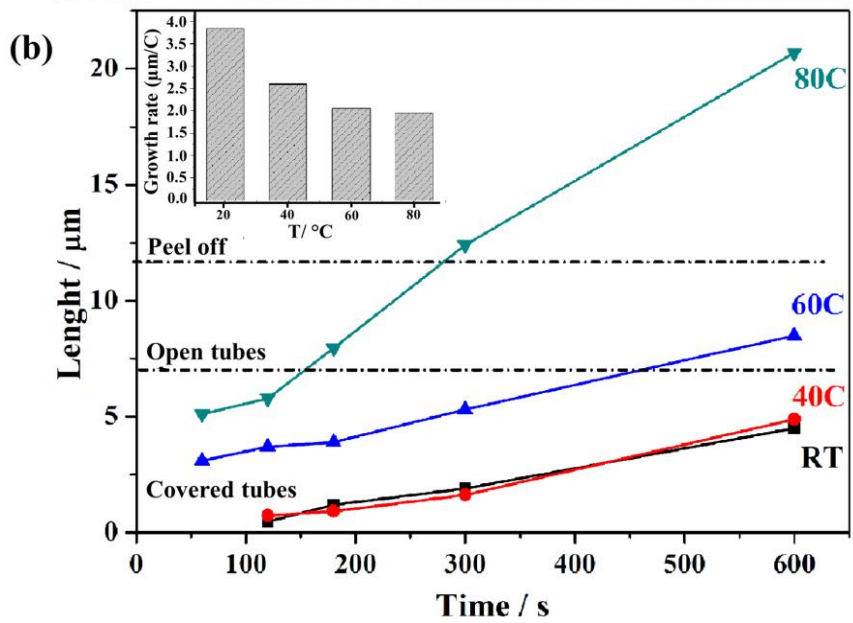

**Figure 3**

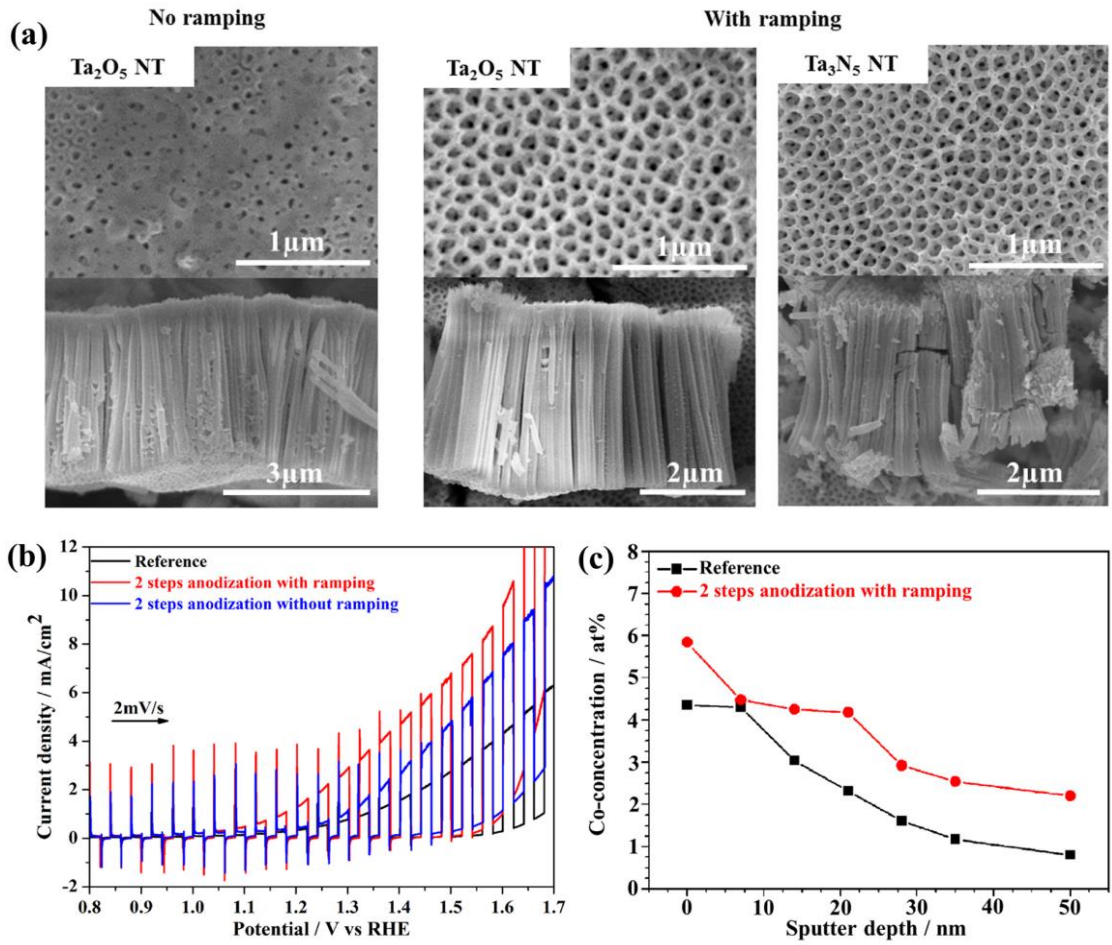

**Figure 4**

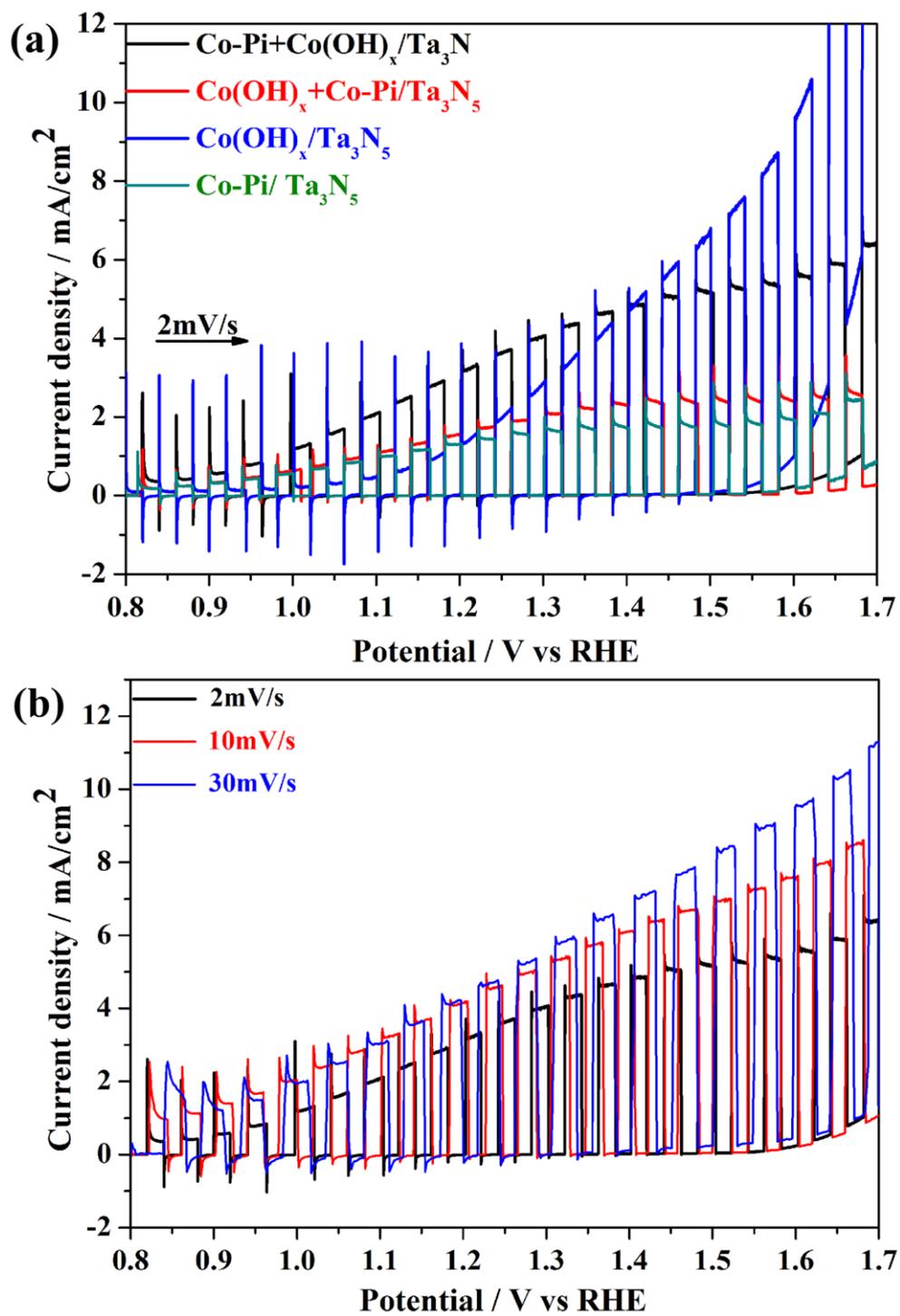